# Trap states impact photon upconversion in rubrene sensitized by lead halide perovskite thin films

*Sarah Wieghold,[1] Alexander S. Bieber,[1] Zachary A. VanOrman,[1] Juan-Pablo Correa-Baena,[2] Lea Nienhaus[1],\**

[1]Department of Chemistry and Biochemistry, Florida State University, Tallahassee, FL 32306, USA
[2]Department of Materials Science and Engineering, Georgia Institute of Technology, Atlanta, GA 30332, USA

*corresponding Author: nienhaus@chem.fsu.edu

**ABSTRACT**

The same optical and electronic properties that make perovskite thin films ideal absorber materials in photovoltaic applications are also beneficial in photon upconversion devices. In this contribution, we investigate the rubrene-triplet sensitization by perovskite thin films based on methylammonium formamidinium lead triiodide (MAFA). To elucidate the role of trap states which affect the free carrier lifetimes, we fabricate MAFA perovskite thin films with three different thicknesses. By measuring the change in the photoluminescence properties under different excitation fluences, we find that the prevalent recombination mechanism shifts from monomolecular for thinner films to bimolecular recombination for thicker MAFA films, indicating a reduction in shallow trap-assisted recombination. The addition of rubrene shows a passivating effect on the MAFA surface, but adds an additional quenching pathway due to charge transfer to the triplet state of rubrene. We observe that the threshold for efficient triplet-triplet annihilation shifts to lower incident powers with increasing MAFA thickness, which suggests that the charge transfer to the triplet state competes with non-radiative trap filling. Hence, injection of free electrons and holes into the upconverting organic semiconductor can provide a new avenue for sensitization of rubrene, and may allow us to move away from the necessity of efficient excitonic singlet-to-triplet converters.

# INTRODUCTION

Photon upconversion (UC) bears the potential in aiding to overcome the Shockley-Queisser[1] limit determining the achievable power conversion efficiency (PCE) of single-junction photovoltaics (PVs) by allowing for the collection of sub-bandgap photons, or to extend the observable wavelengths of silicon-based devices, resulting in low-cost infrared imaging devices.[2–4] In photon UC, two or more low-energy photons are combined to create one higher-energy photon, effectively shortening the wavelength of the light emitted upon irradiation. In organic semiconductors, the UC process is obtained *via* diffusion-mediated triplet-triplet annihilation (TTA)[5,6] in contrast to nonlinear crystals or lanthanide-based nanoparticles, where the process is achieved by second-harmonic frequency generation or through the ladder-like electronic structure of the nanoparticles, respectively.[7] Since the energy is stored in the long-lived triplet states, TTA has the advantage over the other UC processes, that it can become efficient at low photon fluxes, making it applicable at even sub-solar fluxes.[8,9]

In general, TTA-UC systems consist of two parts: a sensitizer which is directly optically excited, and the emitter or annihilator, which is then indirectly excited *via* a spin-allowed Dexter-type triplet energy transfer (TET) process.[10] Thus, one requirement for the sensitizer is the efficient interconversion of singlet states to triplet states, ideally with minimal energy loss in the process. Recently semiconductor nanocrystals (NCs) consisting of lead sulfide (PbS),[9,11–14] cadmium selenide (CdSe)[15] or even $CsPbBr_xI_{1-x}$ perovskites[16] have been employed as triplet sensitizers. Historically metal-organic complexed containing heavy metal atoms were used as sensitizer, however energy losses of up to 300 meV were observed due to the large exchange energies between the singlet and the triplet states.[8,17,18] One drawback of employing NCs as sensitizers are the long insulating ligands passivating the NC surface, which add an additional energy barrier for the TET process and subsequently result in poor exciton transport to the organic-inorganic interface in NC-based device structures. In PbS-based UC devices for example, lack of long-range exciton diffusion restricts the PbS NC layer thickness to one or two monolayers, resulting in very low NIR absorption of under 1%, limiting their achievable external UC efficiency.[11,12,19] To increase the NIR absorption, we seek to replace the NC array with a bulk semiconductor which shows long carrier diffusion lengths and free carrier lifetimes exceeding the characteristic time of energy transfer from the sensitizer to the annihilator. Here, perovskite thin films have shown impressive performances and efficiencies when integrated in perovskite solar cells (PSCs) due to their exceptional material properties. These properties have been attributed to low exciton binding energies,[20–22] low non-radiative recombination,[23] absorption over the visible and near-infrared (NIR) spectrum, and long carrier diffusion lengths due to long carrier lifetimes[24–26] which allow for the efficient extraction of charges at the interfaces, *i.e.* electron- and hole-transport layers. Besides the sensitizer, an annihilator is also necessary in UC devices. Here, rubrene, a p-type semiconductor tetraphenyl derivative of tetracene has been extensively studied due to the exceptional high hole mobilities and chemical stability.[27–29] Recently, rubrene has also been used hole transport layer in PV[30,31] as well as in organic light emitting diodes (OLEDs).[32] Interestingly, the onset of electroluminescence (EL) is observed at ~1eV, or roughly half the rubrene bandgap (2.2 eV) which has been attributed to either Auger-type upconversion,[33] upconversion through a triplet-triplet annihilation (TTA) pathway,[34] or even band-to-band recombination of minority carriers.[35]

To take advantage of the UC process, we investigate the rubrene triplet sensitization by films based on methylammonium (MA), formamidinium (FA) lead triiodide ($MA_{0.15}FA_{0.85}PbI_3$, MAFA) perovskite thin films of varying thicknesses to modify the NIR absorption. The deposited rubrene layer is doped with 1% dibenzotetraphenylperiflanthene (DBP) in a commonly used host-guest/annihilator-emitter approach to increase the quantum yield (QY) of the rubrene film.[14] This addition is required because rubrene is not only capable of TTA but also the reverse process of singlet fission (SF). Förster resonance energy transfer (FRET) of the singlets generated in rubrene by TTA to the emitter DBP outcompetes SF and thus boosts the QY.[14] Rubrene was chosen as the annihilator due to a band alignment known to allow for hole extraction.[36–38] Hence, injection of free electrons and holes into the upconverting organic semiconductor can provide a new avenue for sensitization of rubrene, and may allow us to move away from the necessity of efficient singlet-to-triplet exciton converters.

**RESULTS**

**Absorption and morphology.**

Inspired by the sub-bandgap onset of the rubrene luminescence in OLEDs, we replace the excitonic sensitizer with a bulk perovskite thin film capable of creating long-lived free carriers upon irradiation, effectively using a PV to create highly mobile free charges which can electronically stimulate the rubrene (analogous to an OLED). A major benefit of this approach is the possibility of a higher absorption of the NIR excitation, boosting the UC quantum yield (QY) beyond the limitations of current excitonic UC devices. Since it is known that free carrier lifetimes are highly dependent on the trap state density,[39,40] we fabricate $MA_{0.15}FA_{0.85}PbI_3$ (MAFA) perovskite thin films with three different thicknesses. We have previously shown that by changing the thickness of the perovskite thin films,[41] we can change their PL behavior due to a different number of trap states.[42,43] Here, we fabricate MAFA thin films by changing the molar precursor concentration (dilution factors of 2X, 5X and 10X) resulting in estimated film thicknesses of 200, 45, and 25 nm, respectively. We label the different thicknesses throughout the manuscript as 2X MAFA, 5X MAFA, and 10X MAFA. To investigate the surface morphology of the different MAFA film thicknesses, we perform atomic force microscopy (AFM). Figure S1 details the film morphologies obtained for the 2X, 5X and 10X MAFA films, respectively. A clear change in the morphology is observed based on the different thicknesses. The 10X MAFA film shows a very rough surface with large agglomerates, while the 5X MAFA film shows distinct, yet discontinuous grains. Root mean square (RMS) values of approximately 33 nm and 7 nm can be extracted for the 10X and 5X MAFA films, respectively. We also observe a change in the grain size resulting in slightly bigger grains for the 10X MAFA film. We attribute the bigger grains to remnant domains of $PbI_2$ which result during the growth of perovskite thin films.[44] Further increasing the thickness, yields a smooth, thin film structure for the 2X MAFA concentration with an extracted RMS value of 8 nm. Here, the grain size increases towards an approximately pixel area of 160 $px^2$ (which corresponds to a grain size diameter of ~120 nm by assuming round shaped islands.) The main findings of the AFM measurement are summarized in Figure 1a.

Figure 1b shows the film thickness-dependent absorption spectra of the MAFA films as dashed lines. As expected, the observed optical density increases with the MAFA film thickness, and all films show broadband absorption with an absorption onset at ~800 nm, in line with the expected optical bandgap of 1.55 eV for this perovskite composition.[45]

To investigate the influence of rubrene, we fabricate bilayer devices consisting of MAFA and rubrene/1%DBP (which we label here: MAFArubr). The corresponding absorption spectra are shown in Figure 1b as continuous lines for all three MAFA thicknesses. The overall MAFA absorption decreases in the presence of rubrene, which indicates the solution-based rubrene deposition may slightly affect the underlying MAFA film thickness. The inset in Figure 1b highlights the absorption onset at 800 nm, independent of the film thickness. The characteristic absorption features of the spin-cast rubrene on top of the MAFA films can be clearly seen in the range 430 – 530 nm.[46]

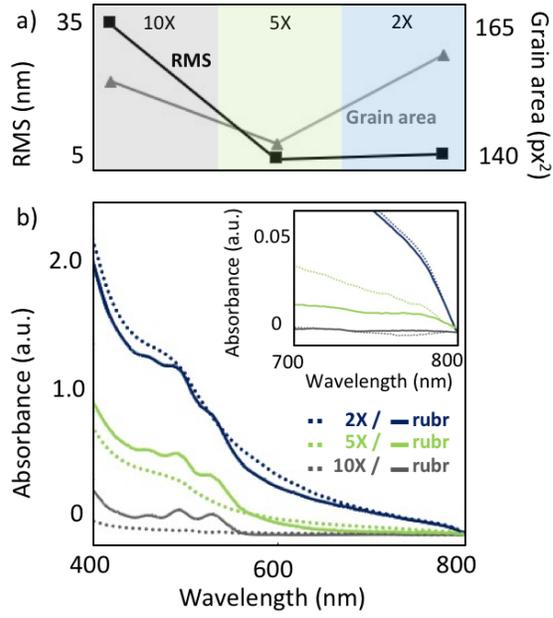

Figure 1: a) RMS values and grain area in px² for the 10X, 5X and 2X MAFA thin films. b) Absorption spectra of the 2X, 5X and 10X MAFA thin films (dashed lines) and the respective MAFArubr bilayer devices highlighting the additional absorption caused by rubrene at 430 – 530 nm (solid lines). The inset shows the absorption onset of the devices at 800 nm, as expected for the 1.55 eV bandgap for the MAFA composition used.

**Power-dependence under CW excitation.**

To study the change in recombination behavior upon increasing the MAFA film thickness, we turn to photoluminescence (PL) spectroscopy. In the first step, we investigate the power dependency of the PL intensity under continuous wave (CW) excitation at 780 nm. Bimolecular free carrier recombination and monomolecular shallow trap-assisted show distinct signatures in their power dependencies, as these are second-order (quadratic) and first-order (linear) processes, respectively. As a result, the power-law dependence of the PL intensity as a function of the excitation density should show a slope $\alpha=2$ for purely free carrier recombination and $\alpha=1$ for trap-assisted recombination. Slopes of $1<\alpha<2$ indicate a mixture of both processes. Figure 2a shows the power-law dependencies of the 2X, 5X and 10X MAFA films on a log-log scale. We observe a change in the slope from α=1.23 for the thinnest 10X film to a free carrier recombination-dominated slope α=2 for the thickest film. It was previously reported, that the addition of rubrene passivates shallow traps at the perovskite surface due to non-covalent cation-π interactions between the $MA^+$ cations and the extended delocalized π-system in rubrene.[37,38] This chelation-like interaction is able to immobilize the otherwise mobile organic cations, which results in a reduction of defect sites during device fabrication, as well as a passivation effect on existing defect sites such as dangling bonds, thus increasing the overall device performance. To confirm this effect, we investigate the power dependencies of the PL of the bilayer upconversion devices. Indeed, we observe a change in the slopes to higher values compared to the MAFA films without rubrene. Figure 2b highlights the power-law dependencies of the 2X, 5X and 10X bilayer MAFArubr devices. The addition of rubrene increases the slope to α=1.42 for the 10X MAFArubr device, and to α=1.68 for the 5X MAFArubr device, further confirming a reduction in trap-assisted recombination in the bilayer devices due to the passivation by rubrene.

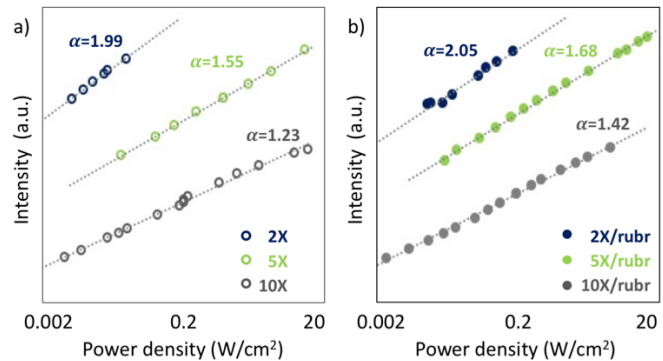

**Figure 2:** a) log-log plot of the NIR PL intensity as a function of the incident excitation power density for the 2X (blue), 5X (green) and 10X (gray) MAFA thin films under 780 nm excitation. The dashed lines are fitted curves to extract the slope α, which increases with increasing film thickness. b) Corresponding power dependence of the MAFArubr bilayer devices. The extracted slopes α shift to higher values, indicating a higher percentage of bimolecular free carrier recombination resulting from defect passivation by rubrene. For a better comparison, the PL intensity traces are offset on the y-axis.

**Time-resolved PL spectroscopy.**

Time-resolved PL (TRPL) spectroscopy is able to yield additional insight on the rate of trap filling, as well as give a qualitative view of the availability of trap states at a given excitation fluence. At low excitation fluences, the light intensity is not sufficient to create enough carriers in the material to fill all existing traps.[47] As a result, the lifetimes show two components under these conditions: a fast component attributed to rapid non-radiative quenching into trap states, as well as a longer component which is then assigned to the free carrier recombination and reflects the carrier lifetime. With increasing incident power intensity, the number of initially empty trap states is reduced, which is reflected in a decreased amplitude of the early time PL quenching component of the lifetime. An increase in free carriers created at higher fluences will manifest as a reduction in the free carrier lifetime, as the likelihood of recombination increases with increasing carrier density. By measuring the change in the PL lifetime under different excitation powers, we are able to extract the rate at which the trap states are populated following optical excitation. Figure 3a shows the power dependent lifetimes of the 2X MAFA film under varying excitation densities at an excitation wavelength of 780 nm. We observe a highly multiexponential decay at a low excitation power (or at low carrier density), while a nearly monoexponential decay is found at high excitation powers. The 5X MAFA film exhibits the same general type of behavior compared the 2X MAFA film (SI Figure S2). However, by further decreasing the film thickness, our 10X MAFA film shows highly multiexponential decays even at the highest available power density. We note, that similar average excitation intensities were used for all three film thicknesses. However, since the film thickness decreases for the 5X and 10X films resulting in different absorption cross sections, a lower excitation density/cm$^3$ is expected for our thinner films at the same incident power. Hence, we may not be able to fully saturate the existing trap states with our available laser power and we still observe early time quenching resulting from trap filling in the thinner films even at the highest fluences available.

The power dependent lifetimes of the 2X MAFArubr device are shown in Figure 3b (see Figure S2 for the 5X and 10X MAFArubr devices). We observe that the addition of rubrene adds a new decay pathway for the generated carriers, which results in an overall shorter, or more quenched lifetime. The same general trend as in the MAFA devices can be seen, where the amount of early time quenching is reduced at higher incident power and the free carrier lifetime is reduced with increasing carrier density. This indicates that the underlying native trap-filling processes still occur in the presence of rubrene, and that the charge extraction to rubrene is likely a competing mechanism. Figure 3c details a schematic of our proposed charge extraction mechanism, where non-radiative trap-filling and charge extraction to rubrene are competing PL quenching pathways. Here, charge extraction to rubrene is enabled because holes can freely transfer from the MAFA perovskite films to rubrene due to a band alignment which readily allows for hole extraction (valence band (VB) of MAFA ~5.8 eV,[45] highest occupied molecular orbital (HOMO) of rubrene ~5.4 eV[11,48]). The large energy mismatch on the order of ~1 eV between the MAFA perovskite conduction band (CB) and the lowest unoccupied molecular orbital (LUMO) creates a barrier for the direct injection of electrons into the singlet excited state of rubrene. However, the bound triplet state $T_1$ may be able to be populated through a charge-transfer (CT) state at the MAFA/rubrene interface. For charge extraction to the rubrene triplet to occur efficiently, the charge transfer must either outcompete the inherent trap-filling process, or sufficient carrier density must be created in the MAFA films to saturate the traps. For the case of lower fluences, there will be a distribution, *i.e.* a probability density function of the carriers trapped *vs.* transferred to rubrene based on the respective rates.

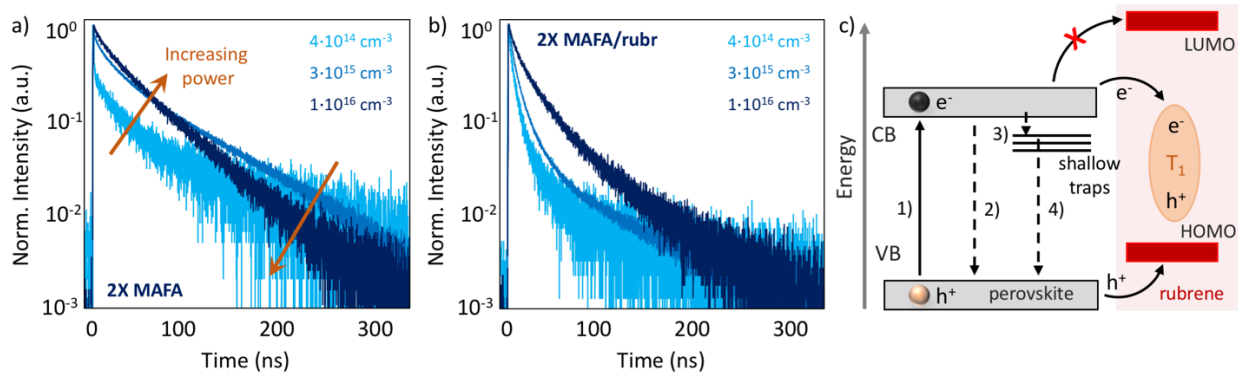

Figure 3: a) NIR PL lifetimes of the 2X MAFA thin film under varying incident powers or carrier densities at an excitation wavelength of 780 nm. The early time quenching diminishes due to an increased amount of trap filling, while the free carrier lifetimes decrease due to an increase in probability of recombination with carrier density. b) NIR PL lifetimes of the 2X MAFArubr bilayer device under the same incident power. An additional early time quenching is observed due to carrier extraction to rubrene. c) Schematic of the proposed rubrene sensitization mechanism: 1) incident light promotes an electron from the VB (~5.8 eV) to the CB (~4.25 eV) of the perovskite. This excitation can be quenched by several pathways: 2) bimolecular free carrier recombination, 3) defect level trapping and 4) trap-assisted recombination or carrier extraction to rubrene. The hole can be readily extracted to the HOMO (~5.4 eV) or rubrene, while the 1 eV mismatch of the perovskite CB and rubrene LUMO blocks direct electron injection into rubrene. However, the bound triplet state $T_1$ of rubrene can be populated through a CT state at the interface.

**Extraction of charge transfer rate.**

We use the approach developed by Wu *et al.*[14] for extracting the additional quenching component caused by charge transfer to rubrene. To obtain the characteristic time of transfer, a scaled copy of the residual sensitizer lifetime is subtracted from the lifetime of the sensitizer in presence of the annihilator and the obtained lifetime is attributed to the energy transfer. Figures 4a, b and c detail the extracted transfer rates for the 2X, 5X and 10X devices, respectively. We observe a clear increase in the extracted transfer rate with increasing excitation power, or carrier density. However, the charge extraction to rubrene should not show a power dependent change in the transfer rate, unless the energy levels involved in the charge transfer, the wavefunction overlap or the attempt frequency are changing, which is not anticipated.

By critically reviewing the extraction approach, we find that one of the underlying assumptions is that the inherent sensitizer decay dynamics are not affected by the addition of rubrene. However, due to the passivating effect of rubrene, the MAFArubr bilayer device will likely saturate its trap states at a lower excitation fluence than the MAFA device, resulting in a different native lifetime at the same excitation power until the excitation density is high enough to saturate the trap states.

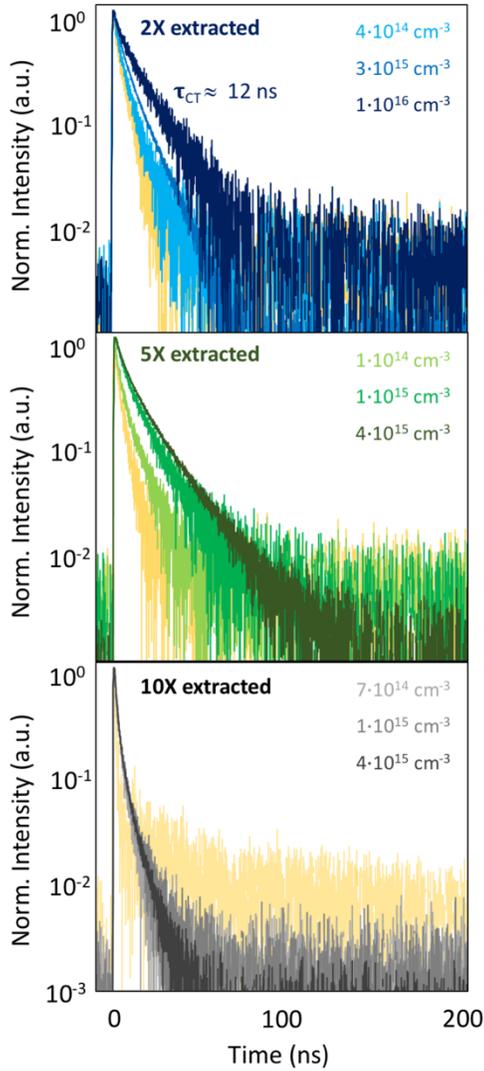

**Figure 4:** Extracted lifetimes of the charge transfer to rubrene for the 2X (top, blue), 5X (middle, green) and 10X (bottom, gray) devices at varying carrier densities (incident power). For the 2X and 5X film, which show the least amount of trap-assisted recombination, we observe an increase in the extracted time of the charge transfer stemming from a fluence-dependent superposition of quenching caused by trap-filling and charge transfer. For the 2X film, at the highest carrier density ($3\cdot10^{16}$ cm$^{-3}$), the characteristic time of transfer is $\tau_{CT} \approx 12$ ns. The 10X device does not show a change in the extracted transfer rate, indicating that even at the highest fluences possible, the MAFArubr PL quenching is dominated by a single pathway: filling of shallow traps. The extracted rate of trap-filling is highlighted in yellow for each device thickness.

Hence, this approach must be taken with a grain of salt. In our case, the fluence dependent change in the characteristic time of transfer shows that the required assumptions do not hold in the case of our devices except at high fluences where the traps are already nearly saturated upon excitation.

Rather, the extraction at low fluences (low carrier densities) yields a superposition of quenching resulting from trap filling and charge transfer to rubrene. To highlight this, we have also included the extracted rate of trap filling for each bilayer device in Figure 4 (yellow). The rate of charge transfer can therefore only be reliably extracted for the 2X and 5X film, which show sufficient carrier density to adequately saturate the existing traps, and we obtain a characteristic time of charge transfer of $\tau_{CT} \approx 12$ ns for the highest incident photon fluence.

**Upconversion process.**

Thus far, we have been able to show that the addition of rubrene passivates the MAFA surface, and that additional quenching of the MAFA PL is observed in the presence of rubrene for our MAFArubr bilayer devices. This is in line with what is expected, as rubrene has been previously investigated as a hole transport layer.[30,31] However, the quenching of the PL does not present clear proof that the electron transfer required for triplet sensitization is also occurring.

To shed more light on the upconversion process, we investigate the blue-shifted emission obtained from the devices under 780 nm excitation under various incident power densities. The upconverted light emission herein relies on three steps: i) charge transfer to the triplet state ii) diffusion-mediated triplet-triplet annihilation (TTA) in rubrene and iii) emission from dibenzotetraphenylperiflanthene (DBP) following Förster resonance energy transfer (FRET). Hence, if upconversion is occurring, emission of the upconverted light is expected at a wavelength of $\lambda \approx 610$ nm, the emission wavelength of our dopant DBP.

Bimolecular TTA has a unique power dependency, with a slope change from $\alpha=2$ at low excitation powers to $\alpha=1$ above the threshold at which TTA becomes efficient.[49] This is due to a change in the underlying kinetics: at low excitation powers triplets decay *via* quasi first-order kinetics and the upconverted PL intensity increases quadratically with excitation power (weak annihilation regime). Above the threshold at which TTA becomes efficient, triplets decay primarily through bimolecular TTA, yielding a linear dependence of the upconverted PL based on the incident power (strong annihilation limit). Figure 5a shows the power dependency of the upconverted light (λ<650 nm) for the bilayer devices with different MAFA thicknesses. All devices exhibit the slope change typical of TTA, indicating that the detected visible PL indeed stems from triplet sensitization of the rubrene layer. We observe that the threshold shifts to lower incident powers with increasing MAFA thickness, which is in line with our proposed mechanism, in which the charge transfer to the triplet state competes with non-radiative trap filling. Thresholds of 300, 700 and 800 mW/cm$^2$ can be extracted for the 2X, 5X, and 10X films, respectively. Because the trap states of our thickest MAFA film are saturated at a lower incident power compared to the thinner films, a more efficient charge transfer to rubrene is expected. These observations are also supported by our observed shift from more "trap-assisted" monomolecular recombination for the thinnest film (10X) to predominately bimolecular recombination in the 2X film for the MAFA emission detailed previously. We note, that no complete saturation of trap states was seen for our 10X MAFA/rubr device under pulsed excitation, and that charge transfer to rubrene likely competes with trap filling throughout the available excitation powers.

In Figure 5b, we show the TRPL lifetime of the upconverted emission. Since triplets are spin-forbidden, and therefore not able to radiatively recombine to the ground state, they exhibit very long lifetimes with reported values up to 100 μs.[50] However, our PL system is limited to a repetition rate of 32.5 kHz, or a detectable time frame of ~33 μs. It has been shown previously, that a too high repetition rate results in an apparent reduction of the triplet lifetime,[11,14] which leads us to be able to only extract a lower bound of the triplet lifetime of ~15 μs in our bilayer devices. The lifetime shows the rise and fall characteristic of the upconversion process, as the observed visible PL is a result of the delayed triplet population by charge transfer and subsequential diffusion-mediated TTA. The upconverted population peaks after ~1.5 µs, which highlights the underlying slow diffusion-mediated process. The initial fast decay observed in the first ~100 ns can be attributed to a small amount of residual MAFA emission not fully removed by the filters.

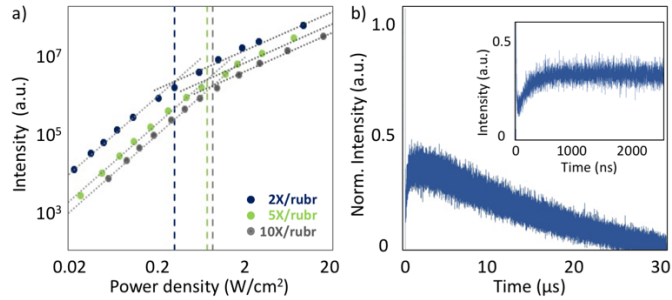

Figure 5: a) Power dependency of the upconverted PL ((λ<650 nm) for the 2X, 5X and 10X MAFArubr bilayer devices. The vertical lines indicate the threshold at which the slope changes from α=2 to α=1 for each of the film thicknesses. b) Dynamics of the upconverted emission. The early time dynamics show an initial rise which peaks at 1.5 µs, and indicates the time required for triplet sensitization and diffusion-mediated TTA to occur. The population then slowly decreases, limited by the long lifetime of the triplet $\tau_{triplet}$ > 15 µs. The inset shows an enlargement of the early time dynamics.

In conclusion, we have been able to investigate the mechanism of triplet sensitization in rubrene by bulk MAFA films. Our results indicate that the inherent traps in the perovskite are filled on a similar timescale as the charge transfer to rubrene occurs, and that both processes compete until the incident power is high enough to fill existing traps. Based on the passivating effect of rubrene caused by cation-π interactions, care must be taken when attempting to extract an accurate rate for the triplet sensitization process as the underlying PL lifetimes are affected by the trap states. Only at high fluences, when existing traps are already filled, can we extract a meaningful rate of charge transfer to the rubrene triplet state: $\tau_{CT} \approx 12$ ns. Increased absorption and reduced defects in a thicker MAFA sensitizer layer result in a reduction of the threshold power density at which the upconversion becomes efficient, paving a clear path to further increase the upconverted PL. Optimization of the charge extraction by tuning the energetic band alignment through dopants in the MAFA film, MAFA defect reduction, and improving the device structure provide a means to further increase the device performance.

**Materials and Methods**

**Device Fabrication**

The perovskite films were prepared according using PbI2 (1.2 M, 99.99% Aldrich) and MAI (1.2 M, Dyenamo), in a 1:1 molar ratio, and PbI2 (1.2 M, 99.99% Aldrich) and FAI (1.2 M, Dyenamo), in a 1:1 molar ratio, both dissolved in anhydrous DMF:DMSO 9:1 (v:v, Acros).[51] The perovskite precursor solutions were then diluted 2-fold (2X), 5-fold (5X) and 10-fold (10X) to fabricate the different film thicknesses.

Glass substrates were cleaned with acetone and then placed in a UV ozone plasma cleaner for 10 min. The following two-step program was used to spin-coat the perovskite; first at 1000 rpm for 10 s and then at 4000 rpm for 30 s. During the second spin coating step, 100 μL of chlorobenzene were dropped on the substrate 20 seconds before the end of the program. The substrates were then annealed at 100 °C for 5 min under nitrogen atmosphere.

The rubrene/DBP solution was prepared in toluene (Sigma Aldrich) at 10 mg/mL. The rubrene solution was doped with DBP at a 1% molar ratio from a 1 mg/mL stock solution. The solution was deposited by spin coating on the perovskite layers at 6000 rpm for 20 seconds. In order to avoid contact with oxygen and moisture with the device, the upconversion devices were sealed using a 2-part epoxy inside the glovebox.

**Atomic Force Microscopy**

AFM images were taken using an Asylum MFP-3D ambient AFM in tapping mode, with a silicon cantilever (300 kHz, spring constant: 26 N/m).

**Absorption spectroscopy**

Absorption spectra were measured using a Shimadzu UV-2450 UV-Vis spectrometer.

**Photoluminescence spectroscopy.**

Time-resolved photoluminescent (TRPL) lifetimes were obtained by time-correlated single photon counting (TCSPC). All samples were excited by a picosecond pulsed diode laser (PicoQuant LDH-D-C-780) connected to a PicoQuant Laser Driver (PDL 800-D). Excitation powers were measured using a silicon power meter (ThorLabs PM100-D). Upconverted lifetimes were obtained at an average excitation power of 3.1 μW and a repetition rate of 32.5 kHz. Excess laser scatter and perovskite photoluminescent emission were removed by 650 nm and 700 nm shortpass filters (ThorLabs). To monitor the perovskite emission, lifetimes were obtained at various excitation power densities at a repetition rate of 2.5 MHz. Laser scatter was removed with an 800 nm longpass filter (ThorLabs). For the power dependent PL intensities, the 780 nm laser was used in CW-mode for a 20 s histogram time. In all cases, the resulting emission was focused onto a single photon counting avalanche photodiode (MPD). A HydraHarp 400 (PicoQuant) was used to record the photon arrival times.

The laser spot size was determined by the razor blade method, yielding a spot size of 150 μm based on the $1/e^2$ distance.

**AUTHOR INFORMATION**

**Corresponding Author:**

*Lea Nienhaus: nienhaus@chem.fsu.edu

ACKNOWLEDGEMENT

The authors would like to thank the Mattoussi group at FSU for aiding in the UV-Vis measurements and gratefully acknowledge Florida State University startup funds.

**REFERENCES**
(1) Shockley, W.; Queisser, H. J. Detailed Balance Limit of Efficiency of p-n Junction Solar Cells. *J. Appl. Phys.* **1961**, *32*, 510–519.
(2) Trupke, T.; Green, M. A.; Würfel, P. Improving Solar Cell Efficiencies by Up-Conversion of Sub-Band-Gap Light. *J. Appl. Phys.* **2002**, *92*, 4117–4122.
(3) Meng, F.-L.; Wu, J.-J.; Zhao, E.-F.; Zheng, Y.-Z.; Huang, M.-L.; Dai, L.-M.; Tao, X.; Chen, J.-F. High-Efficiency near-Infrared Enabled Planar Perovskite Solar Cells by Embedding Upconversion Nanocrystals. *Nanoscale* **2017**, *9*, 18535–18545.
(4) He, M.; Pang, X.; Liu, X.; Jiang, B.; He, Y.; Snaith, H.; Lin, Z. Monodisperse Dual-Functional Upconversion Nanoparticles Enabled Near-Infrared Organolead Halide Perovskite Solar Cells. *Angewan. Chem.* **2016**, *128*, 4352–4356.
(5) Schmidt, T. W.; Castellano, F. N. Photochemical Upconversion: The Primacy of Kinetics. *J. Phys. Chem. Lett.* **2014**, *5*, 4062–4072.
(6) Schulze, T. F.; Schmidt, T. W. Photochemical Upconversion: Present Status and Prospects for Its Application to Solar Energy Conversion. *Energy Environ. Sci.* **2014**, *8* (1), 103–125.
(7) Zhou, J.; Liu, Q.; Feng, W.; Sun, Y.; Li, F. Upconversion Luminescent Materials: Advances and Applications. *Chem. Rev.* **2015**, *115*, 395–465.
(8) Singh-Rachford, T. N.; Castellano, F. N. Photon Upconversion Based on Sensitized Triplet–Triplet Annihilation. *Coord. Chem. Revs.* **2010**, *254*, 2560–2573. https://doi.org/10.1016/j.ccr.2010.01.003.
(9) Mahboub, M.; Huang, Z.; Tang, M. L. Efficient Infrared-to-Visible Upconversion with Subsolar Irradiance. *Nano Lett.* **2016**, *16*, 7169–7175.
(10) Dexter, D. L. A Theory of Sensitized Luminescence in Solids. *J. Chem. Phys.* **1953**, *21*, 836–850.
(11) Nienhaus, L.; Wu, M.; Geva, N.; Shepherd, J. J.; Wilson, M. W. B.; Bulović, V.; Van Voorhis, T.; Baldo, M. A.; Bawendi, M. G. Speed Limit for Triplet-Exciton Transfer in Solid-State PbS Nanocrystal-Sensitized Photon Upconversion. *ACS Nano* **2017**, *11*, 7848–7857.
(12) Nienhaus, L.; Wu, M.; Bulovic, V.; Baldo, M. A.; Bawendi, M. G. Using Lead Chalcogenide Nanocrystals as Spin Mixers: A Perspective on near-Infrared-to-Visible Upconversion. *Dalton Trans.* **2018**, *47*, 8509–8516.
(13) Huang, Z.; Li, X.; Mahboub, M.; Hanson, K. M.; Nichols, V. M.; Le, H.; Tang, M. L.; Bardeen, C. J. Hybrid Molecule–Nanocrystal Photon Upconversion Across the Visible and Near-Infrared. *Nano Lett.* **2015**, *15*, 5552–5557.
(14) Wu, M.; Congreve, D. N.; Wilson, M. W. B.; Jean, J.; Geva, N.; Welborn, M.; Van Voorhis, T.; Bulović, V.; Bawendi, M. G.; Baldo, M. A. Solid-State Infrared-to-Visible Upconversion Sensitized by Colloidal Nanocrystals. *Nature Photon.* **2016**, *10*, 31–34.

(15) Huang, Z.; Li, X.; Yip, B. D.; Rubalcava, J. M.; Bardeen, C. J.; Tang, M. L. Nanocrystal Size and Quantum Yield in the Upconversion of Green to Violet Light with CdSe and Anthracene Derivatives. *Chem. Mater.* **2015**, *27*, 7503–7507.

(16) Mase, K.; Okumura, K.; Yanai, N.; Kimizuka, N. Triplet Sensitization by Perovskite Nanocrystals for Photon Upconversion. *Chem. Commun.* **2017**, *53*, 8261–8264.

(17) Monguzzi, A.; Tubino, R.; Meinardi, F. Multicomponent Polymeric Film for Red to Green Low Power Sensitized Up-Conversion. *J. Phys. Chem.* **2009**, *113*, 1171–1174.

(18) Amemori, S.; Sasaki, Y.; Yanai, N.; Kimizuka, N. Near-Infrared-to-Visible Photon Upconversion Sensitized by a Metal Complex with Spin-Forbidden yet Strong $S_0-T_1$ Absorption. *J. Am. Chem. Soc.* **2016**, *138*, 8702–8705.

(19) Nienhaus, L.; Geva, N.; Correa-Baena, J.; Wu, M.; Wieghold, S.; Bulović, V.; Voorhis, T. V.; Baldo, M. A.; Buonassisi, T.; Bawendi, M. G. Solid-State Infrared-to-Visible Upconversion for Sub-Bandgap Sensitization of Photovoltaics. *2018 IEEE 7th World Conference on Photovoltaic Energy Conversion (WCPEC) (A Joint Conference of 45th IEEE PVSC, 28th PVSEC 34th EU PVSEC)* **2018**, 3698–3702.

(20) Galkowski, K.; Mitioglu, A.; Miyata, A.; Plochocka, P.; Portugall, O.; Eperon, G. E.; Wang, J. T.-W.; Stergiopoulos, T.; Stranks, S. D.; Snaith, H. J.; et al. Determination of the Exciton Binding Energy and Effective Masses for Methylammonium and Formamidinium Lead Tri-Halide Perovskite Semiconductors. *Energy Environ. Sci.* **2016**, *9*, 962–970.

(21) Yang, Y.; Yang, M.; Li, Z.; Crisp, R.; Zhu, K.; Beard, M. C. Comparison of Recombination Dynamics in $CH_3NH_3PbBr_3$ and $CH_3NH_3PbI_3$ Perovskite Films: Influence of Exciton Binding Energy. *J. Phys. Chem. Lett.* **2015**, *6*, 4688–4692.

(22) Yamada, Y.; Nakamura, T.; Endo, M.; Wakamiya, A.; Kanemitsu, Y. Photocarrier Recombination Dynamics in Perovskite $CH_3NH_3PbI_3$ for Solar Cell Applications. *J. Am. Chem. Soc.* **2014**, *136*, 11610-11613.

(23) Tress, W. Perovskite Solar Cells on the Way to Their Radiative Efficiency Limit – Insights Into a Success Story of High Open-Circuit Voltage and Low Recombination. *Adv. Energy Mater.* **2017**, *7*, 1602358.

(24) Stranks, S. D.; Eperon, G. E.; Grancini, G.; Menelaou, C.; Alcocer, M. J. P.; Leijtens, T.; Herz, L. M.; Petrozza, A.; Snaith, H. J. Electron-Hole Diffusion Lengths Exceeding 1 Micrometer in an Organometal Trihalide Perovskite Absorber. *Science* **2013**, *342*, 341–344.

(25) Xing, G.; Mathews, N.; Sun, S.; Lim, S. S.; Lam, Y. M.; Grätzel, M.; Mhaisalkar, S.; Sum, T. C. Long-Range Balanced Electron- and Hole-Transport Lengths in Organic-Inorganic $CH_3NH_3PbI_3$. *Science* **2013**, *342*, 344.

(26) Tress, W.; Marinova, N.; Inganäs, O.; Nazeeruddin, M. K.; Zakeeruddin, S. M.; Graetzel, M. Predicting the Open-Circuit Voltage of $CH_3NH_3PbI_3$ Perovskite Solar Cells Using Electroluminescence and Photovoltaic Quantum Efficiency Spectra: The Role of Radiative and Non-Radiative Recombination. *Adv. Energy Mater.* **2014**, *5*, 1400812.

(27) Currie, M. J.; Mapel, J. K.; Heidel, T. D.; Goffri, S.; Baldo, M. A. High-Efficiency Organic Solar Concentrators for Photovoltaics. *Science* **2008**, *321*, 226.

(28) Sundar, V. C.; Zaumseil, J.; Podzorov, V.; Menard, E.; Willett, R. L.; Someya, T.; Gershenson, M. E.; Rogers, J. A. Elastomeric Transistor Stamps: Reversible Probing of Charge Transport in Organic Crystals. *Science* **2004**, *303*, 1644.

(29) Podzorov, V.; Menard, E.; Borissov, A.; Kiryukhin, V.; Rogers, J. A.; Gershenson, M. E. Intrinsic Charge Transport on the Surface of Organic Semiconductors. *Phys. Rev. Lett.* **2004**, *93*, 086602.

(30) Tavakoli, M. M.; Tavakoli, R.; Prochowicz, D.; Yadav, P.; Saliba, M. Surface Modification of a Hole Transporting Layer for an Efficient Perovskite Solar Cell with an Enhanced Fill Factor and Stability. *Mol. Syst. Des. Eng.* **2018**, *3*, 717–722.

(31) Ji, G.; Zheng, G.; Zhao, B.; Song, F.; Zhang, X.; Shen, K.; Yang, Y.; Xiong, Y.; Gao, X.; Cao, L.; et al. Interfacial Electronic Structures Revealed at the Rubrene/CH3NH3PbI3 Interface. *Phys. Chem. Chem. Phys.* **2017**, *19*, 6546–6553.

(32) Okumoto, K.; Kanno, H.; Hamada, Y.; Takahashi, H.; Shibata, K. High Efficiency Red Organic Light-Emitting Devices Using Tetraphenyldibenzoperiflanthene-Doped Rubrene as an Emitting Layer. *Appl. Phys. Lett.* **2006**, *89*, 013502.

(33) Pandey, A. K.; Nunzi, J.-M. Upconversion Injection in Rubrene/Perylene-Diimide-Heterostructure Electroluminescent Diodes. *Appl. Phys. Lett.* **2007**, *90* (26), 263508.

(34) Pandey, A. K. Highly efficient spin-conversion effect leading to energy up-converted electroluminescence in singlet fission photovoltaics, *Sci. Rep.* **2015**, *5*, 7787.

(35) Engmann, S.; Barito, A. J.; Bittle, E. G.; Giebink, N. C.; Richter, L. J.; Gundlach, D. J. Higher Order Effects in Organic LEDs with Sub-Bandgap Turn-On. *Nature Commun.* **2019**, *10*, 227.

(36) Cong, S.; Yang, H.; Lou, Y.; Han, L.; Yi, Q.; Wang, H.; Sun, Y.; Zou, G. Organic Small Molecule as the Underlayer Toward High Performance Planar Perovskite Solar Cells. *ACS Appl. Mater. Interfaces* **2017**, *9*, 2295–2300.

(37) Qin, P.; Zhang, J.; Yang, G.; Yu, X.; Li, G. Potassium-Intercalated Rubrene as a Dual-Functional Passivation Agent for High Efficiency Perovskite Solar Cells. *J. Mater. Chem. A,* **2019**, *7*, 1824-1834.

(38) Wei, D.; Ma, F.; Wang, R.; Dou, S.; Cui, P.; Huang, H.; Ji, J.; Jia, E.; Jia, X.; Sajid, S.; et al. Ion-Migration Inhibition by the Cation–π Interaction in Perovskite Materials for Efficient and Stable Perovskite Solar Cells. *Adv. Mater.* **2018**, *30*, 1707583.

(39) Sherkar, T. S.; Momblona, C.; Gil-Escrig, L.; Ávila, J.; Sessolo, M.; Bolink, H. J.; Koster, L. J. A. Recombination in Perovskite Solar Cells: Significance of Grain Boundaries, Interface Traps, and Defect Ions. *ACS Energy Lett.* **2017**, *2*, 1214–1222.

(40) Chu, Z.; Yang, M.; Schulz, P.; Wu, D.; Ma, X.; Seifert, E.; Sun, L.; Li, X.; Zhu, K.; Lai, K. Impact of Grain Boundaries on Efficiency and Stability of Organic-Inorganic Trihalide Perovskites. *Nature Commun.* **2017**, *8*, 2230.

(41) Correa-Baena, J.-P.; Anaya, M.; Lozano, G.; Tress, W.; Domanski, K.; Saliba, M.; Matsui, T.; Jacobsson, T. J.; Calvo, M. E.; Abate, A.; et al. Unbroken Perovskite: Interplay of Morphology, Electro-Optical Properties, and Ionic Movement. *Adv. Mater* **2016**, *28*, 5031–5037.

(42) Wieghold, S.; Correa-Baena, J.-P.; Nienhaus, L.; Sun, S.; Shulenberger, K. E.; Liu, Z.; Tresback, J. S.; Shin, S. S.; Bawendi, M. G.; Buonassisi, T. Precursor Concentration Affects Grain Size, Crystal Orientation, and Local Performance in Mixed-Ion Lead Perovskite Solar Cells. *ACS Appl. Energy Mater.* **2018**, *1*, 6801-6808.

(43) S. Wieghold; J. Correa-Baena; L. Nienhaus; S. Sun; J. S. Tresback; Z. Liu; S. S. Shin; M. G. Bawendi; T. Buonassisi. Interplay of Grain Size, Crystal Orientation, and Performance in Mixed-ion Lead Halide Perovskite Films. *2018 IEEE 7th World Conference on Photovoltaic Energy Conversion (WCPEC) (A Joint Conference of 45th IEEE PVSC, 28th PVSEC & 34th EU PVSEC)*, **2018**, 2553–2556.

(44) Jacobsson, T. J.; Correa-Baena, J.-P.; Anaraki, E. H.; Philippe, B.; Stranks, S. D.; Bouduban, M. E. F.; Tress, W.; Schenk, K.; Teuscher, J.; Moser, J.-E.; et al. Unreacted PbI2 as a Double-Edged Sword for Enhancing the Performance of Perovskite Solar Cells. *J. Am. Chem. Soc.* **2016**, *138*, 10331–10343.

(45) Correa Baena, J. P.; Steier, L.; Tress, W.; Saliba, M.; Neutzner, S.; Matsui, T.; Giordano, F.; Jacobsson, T. J.; Srimath Kandada, A. R.; Zakeeruddin, S. M.; et al. Highly Efficient Planar Perovskite Solar Cells through Band Alignment Engineering. *Energy Environ. Sci.* **2015**, *8*, 2928–2934.

(46) Ma, L.; Zhang, K.; Kloc, C.; Sun, H.; Michel-Beyerle, M. E.; Gurzadyan, G. G. Singlet Fission in Rubrene Single Crystal: Direct Observation by Femtosecond Pump–Probe Spectroscopy. *Phys. Chem. Chem. Phys.* **2012**, *14*, 8307–8312.

(47) Kim, J.; Godin, R.; Dimitrov, S. D.; Du, T.; Bryant, D.; McLachlan, M. A.; Durrant, J. R. Excitation Density Dependent Photoluminescence Quenching and Charge Transfer Efficiencies in Hybrid Perovskite/Organic Semiconductor Bilayers. *Adv. Energy Mater.* **2018**, *8*, 1802474.

(48) Ji, G.; Zheng, G.; Zhao, B.; Song, F.; Zhang, X.; Shen, K.; Yang, Y.; Xiong, Y.; Gao, X.; Cao, L.; et al. Interfacial Electronic Structures Revealed at the Rubrene/CH3NH3PbI3 Interface. *Phys. Chem. Chem. Phys.* **2017**, *19*, 6546–6553.

(49) Cheng, Y. Y.; Khoury, T.; Clady, R. G. C. R.; Tayebjee, M. J. Y.; Ekins-Daukes, N. J.; Crossley, M. J.; Schmidt, T. W. On the Efficiency Limit of Triplet–Triplet Annihilation for Photochemical Upconversion. *Phys. Chem. Chem. Phys.* **2010**, *12*, 66–71.

(50) Ryasnyanskiy, A.; Biaggio, I. Triplet Exciton Dynamics in Rubrene Single Crystals. *Phys. Rev. B* **2011**, *84* (19).

(51) Jacobsson, T. J.; Correa-Baena, J.-P.; Pazoki, M.; Saliba, M.; Schenk, K.; Grätzel, M.; Hagfeldt, A. Exploration of the Compositional Space for Mixed Lead Halogen Perovskites for High Efficiency Solar Cells. *Energy Environ. Sci.* **2016**, *9*, 1706–1724.

**TOC**

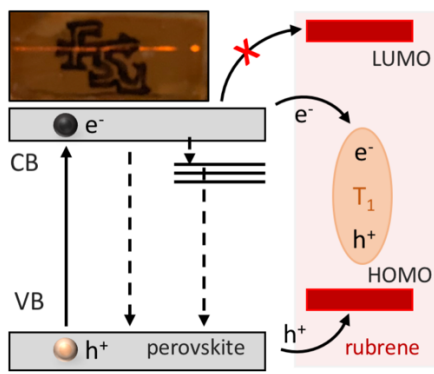